\documentclass[useAMS,usenatbib]{mn2e}
\usepackage{amsmath}
\usepackage{graphicx}
\usepackage{amssymb}
\usepackage{sidecap}
\usepackage{color}
\usepackage{enumerate}
\usepackage{algorithm}
\usepackage{algpseudocode}
\usepackage{verbatim}

\def\dalemb#1#2{{\vbox{\hrule height.#2pt
        \hbox{\vrule width.#2pt height#1pt \kern#1pt \vrule width.#2pt}
        \hrule height.#2pt}}}

\def\ba{\begin{eqnarray}}
\def\ea{\end{eqnarray}}
\def\be{\begin{equation}}
\def\ee{\end{equation}}

\def\gtorder{\mathrel{\raise.3ex\hbox{$>$}\mkern-14mu
             \lower0.6ex\hbox{$\sim$}}}
\def\ltorder{\mathrel{\raise.3ex\hbox{$<$}\mkern-14mu
             \lower0.6ex\hbox{$\sim$}}}

\def\be{\begin{equation}}
\def\ee{\end{equation}}

\def\gtorder{\mathrel{\raise.3ex\hbox{$>$}\mkern-14mu
             \lower0.6ex\hbox{$\sim$}}}
\def\ltorder{\mathrel{\raise.3ex\hbox{$<$}\mkern-14mu
             \lower0.6ex\hbox{$\sim$}}}

\title[Sampling techniques for parameter estimation]{Comparison of sampling techniques for Bayesian parameter estimation}

\author[R.~Allison et al.]{\parbox{\textwidth}{
Rupert~Allison\thanks{E-mail:~\texttt{rupert.allison@astro.ox.ac.uk}},
Joanna~Dunkley}
\vspace{0.4cm}\\
\parbox{\textwidth}{
Sub-department of Astrophysics, University of Oxford, Denys Wilkinson Building, Oxford, OX1 3RH, UK \\
}}

\begin{document}

\date{}

\pagerange{\pageref{firstpage}--\pageref{lastpage}} \pubyear{2013}

\maketitle

\begin{abstract} 
The posterior probability distribution for a set of model parameters encodes all that the data have to tell us in the context of a given model; it is the fundamental quantity for Bayesian parameter estimation. In order to infer the posterior probability distribution we have to decide how to explore parameter space. Here we compare three prescriptions for how parameter space is navigated, discussing their relative merits. We consider Metropolis-Hasting sampling, nested sampling and affine-invariant ensemble MCMC sampling. We focus on their performance on toy-model Gaussian likelihoods and on a real-world cosmological data set. We outline the sampling algorithms themselves and elaborate on performance diagnostics such as convergence time, scope for parallelisation, dimensional scaling, requisite tunings and suitability for non-Gaussian distributions. We find that nested sampling delivers high-fidelity estimates for posterior statistics at low computational cost, and should be adopted in favour of Metropolis-Hastings in many cases. Affine-invariant MCMC is competitive when computing clusters can be utilised for massive parallelisation. Affine-invariant MCMC and existing extensions to nested sampling naturally probe multi-modal and curving distributions. 
\end{abstract}


\label{firstpage}

\section{Introduction}
In the framework of Bayesian data analysis we can rigorously discriminate between competing models and determine the region of parameter space that is favoured by the data. We must be able to traverse parameter space in an efficient manner in order to evaluate both the posterior probability distribution for the parameters and the Bayesian evidence, which is the pertinent quantity for model selection. The interpretation and analysis of empirical data using Bayesian methods have widespread scientific value \citep[e.g.,][]{Gilks:1995, Qian:2003, deAustri:2006, vonToussaint:2011, Parkinson:2013, Mana:2013}.  

The scientific literature is filled with examples of particular sampling techniques prescribing how parameter space is navigated \citep[e.g.,][]{Metropolis:1953, Mackay:2003, Skilling:2004, Feroz:2013}. Many distinct branches of science make use of sampling methods to intelligently traverse the parameter space of their models. Different methods are, of course, tailored to particular problems. In the cosmology community Markov chain Monte Carlo (MCMC) methods are widely used \citep[e.g.,][]{Lewis:2002}. Because MCMC methods are applied ubiquitously, we discuss them in detail here (both single-particle and ensemble methods), comparing them to nested sampling, which was introduced by~\cite{Skilling:2004}. Each method has its own advantages, such as speed, simplicity, range of applicability and scope for parallelisation. Indeed it is the aim of this paper to assess the relative merits of each. Throughout this discussion we will highlight the requirement to minimise computations of the likelihood, subject to returning an accurate approximation to the joint posterior distribution. This is an important diagnostic; in cosmology, likelihood evaluations often involve running a simulation, or solving a set of coupled differential equations. As such the calls to the likelihood function are the limiting factor computationally. It is in this sense that we require a sampler to be efficient.

We begin in \S \ref{ssBayes} by summarising the fundamental results of Bayesian inference, introducing much of the subsequent notation. We then detail the three sampling techniques which are the focus of this paper. Single-particle Metropolis-Hastings sampling is discussed in \S \ref{ssMH}. The formalism and implementation of nested sampling are outlined in \S \ref{ssNS}, while a summary of the Goodman \& Weare affine-invariant ensemble MCMC sampler is given in \S \ref{ssAffInv}. Their relative performance on a toy-model Gaussian likelihood and a realistic cosmological model are tested in \S \ref{sResults} and \S \ref{sCosmoParams}, respectively. We discuss these results in \S \ref{sDiscussion}, and conclude in \S \ref{sConc}.

\section{Exploring parameter space}
\label{sSampling}
\subsection{Bayesian Inference}
\label{ssBayes}
Suppose we have a model $M$ for an observed phenomenon. We would like to determine the posterior probability distribution $\mathbb{P}({\bf \Theta|D},M)$ for the parameters $\bf{\Theta}$ which describe the model, given our knowledge of some data $\bf{D}$ about that phenomenon \citep{Jaynes:1957, Jaynes:2003}. From this quantity we can derive all the usual statistics we are interested in, such as parameter means, uncertainties and correlations. Bayes' theorem allows us to express the posterior as a function of simpler, tractable terms \citep{Bayes:1958}: 
\begin{equation}
\mathbb{P}({\bf \Theta|D},M)=
	\frac{\mathbb{P}({\bf D|\Theta},M)\mathbb{P}({\bf \Theta}|M)}
						{\mathbb{P}({\bf D}|M)}.
\label{bayes}
\end{equation}
Here $\mathcal{L}({\bf \Theta}) \equiv \mathbb{P}({\bf D|\Theta},M)$ is the likelihood, which expresses the explanatory power of a given set of parameters. The prior $\pi({\bf \Theta}) \equiv \mathbb{P}({\bf \Theta}|M)$ represents our degree of belief about the parameters before we have knowledge of the data. We denote the Bayesian evidence $\mathcal{Z} \equiv {\mathbb{P}({\bf D}|M)}$, which for any given model can be written down in integral form,
\begin{equation}
\label{eEvidence}
\mathcal{Z}=\int{\mathcal{L}({\bf \Theta})\pi({\bf \Theta}) d{\bf \Theta}},
\end{equation}
since it is the normalisation factor in Eq.~\ref{bayes}. This formalism allows us to estimate parameters. Now consider the case where we have two competing models $M_1$ and $M_2$, both of which purport to describe the same phenomenon.  Given some data $\bf{D}$ and overall scientific context $I$ (this was suppressed in the above notation), we can again make use of Bayes' theorem to find
\begin{equation}
\frac{\mathbb{P}(M_1|{\bf D},I)}{\mathbb{P}(M_2|{\bf D},I)}=\frac{\mathcal{Z}_1}{\mathcal{Z}_2}
	\frac{\mathbb{P}(M_1|I)}{\mathbb{P}(M_2|I)}.
\label{modelsel}
\end{equation}
Assuming that our {\it a priori} belief in the two models is equal, then the evidence ratio (also called the {\it Bayes factor}) completely specifies the relative probability of the two models \citep[e.g.,][]{Sivia:1996, Hobson:2002}.

The posterior encodes all that the data have to tell us in the context of a given model. Therefore the salient task, for parameter estimation, is to construct an accurate approximation to the posterior. To do this we must sample the posterior sufficiently densely that numerical uncertainties on any quantity we evaluate are negligible for our purposes. Likelihood evaluations can be computationally expensive; for example, calculating the Cosmic Microwave Background (CMB) power spectrum likelihood at one set of cosmological parameters requires us to evolve many coupled differential equations from inflation to the present day, which can take on the order of seconds to evaluate \citep[e.g.,][]{Lewis:1999}. Simple, brute force grid sampling is therefore not feasible since the number of likelihood evaluations scales exponentially with the number of parameters. There are alternatives however, many of which are particular manifestations of the Markov chain Monte Carlo (MCMC) class of sampling algorithms. MCMC samplers work by constructing a Markov chain in parameter space whose equilibrium distribution is the posterior itself. We summarise the implementation of two such algorithms below. We also outline the method of nested sampling, a non-MCMC prescription for how parameter space is explored. 

\subsection{Metropolis-Hastings sampling}
\label{ssMH}
The Metropolis-Hastings algorithm is one of the simplest MCMC sampling methods and has been applied to a huge variety of parameter estimation problems \citep{Metropolis:1953}. We summarise the essentials of the method for a one-dimensional problem with posterior $p(x)$ as follows:
\begin{enumerate}[1.]
\item{Choose initial point in parameter space $x_0$}
\item{At each step $x_i$ propose a trial step $x_\text{trial}$ drawn from a symmetric trial distribution $q(x_\text{trial},x_i)$; this is often taken to be a Gaussian centred on $x_i$.}
\label{step2}
\item{Define $P = \min \left\{1, p(x_\text{trial})/p(x_i) \right\}$. Set $x_{i+1} = x_\text{trial}$ with probability $P$, set $x_{i+1} = x_i$ with probability $1-P$.} 
\item{Iterate from step~\ref{step2} to obtain the chain of points $\left\{x_i\right\}$.}
\end{enumerate}
For more details see e.g., \cite{Lewis:2002} and  \cite{Dunkley:2005}. Thus steps towards regions of higher posterior probability (`uphill steps') are always taken, while downhill steps are taken only occasionally, and are less probable for larger descents. The symmetry of the trial distribution $q$ guarantees the stationarity of $p(x)$ under the Markov process and thus that the asymptotic distribution of the chain is $p(x)$. 

For a $D$-dimensional parameter space in which we have $N$ samples of the posterior $\left\{ \mathbf{x}_i \right\}$, the mean and covariance of the posterior are estimated using
\begin{equation}
\label{eMeanAndCov}
\bar{{\bf x}}=\frac{1}{N}  \sum_{i = 1}^{N} \mathbf{x}_i, \hspace{4  mm}
{\bf C}=\frac{1}{N} \sum_{i = 1}^{N} (\mathbf{x}_i - \bar{{\bf x}})(\mathbf{x}_i - \bar{{\bf x}})^ {\text {\bf T}}.
\end{equation}
Although the Markov chain is guaranteed to converge asymptotically to a perfect sampling of the posterior, we require, in practice, a high-quality sampling in as few chain steps as possible to mitigate computational overheads. This requires a judicious choice of trial distribution, the shape of which can drastically affect the acceptance rate and convergence time \citep[e.g.,][]{Gelman:1996}. For a Gaussian likelihood with covariance $\mathbf{C}_0$ and uniform priors,~\cite{Dunkley:2005} showed the optimal choice for the trial distribution is a Gaussian with covariance 
\begin{equation}
\label{tuning}
\mathbf{C} = (2.4^2/D) \mathbf{C}_0.
\end{equation}
In this case the number of likelihood evaluations for convergence to a given level of accuracy scales linearly with dimension $D$:
\begin{equation}
\label{eMHconv}
N_\text{like} \approx 330D.
\end{equation}
Here convergence is defined by the requirement that the variance of the chain means, $\sigma^2_{\bar x}$, is much less than the variance of the underlying distribution, $\sigma^2_0$, for each parameter $x$ ($\sigma^2_0$ is estimated from the within-chain variance). \cite{Dunkley:2005} define the convergence statistic
\begin{equation}
\label{er}
r = \frac{\sigma^2_{\bar x}}{\sigma^2_0}.
\end{equation}
One can estimate $r$ by spectral methods, or by running several parallel chains and evaluating it directly; we stop the chains when $r<0.01$. The widely used Gelman-Rubin parameter $R$ is related to this statistic by $R \approx 1+r$ \citep{Gelman:1992}.

The cosmology community has historically favoured Metropolis-Hastings as a parameter space sampler \citep[e.g.,][]{Spergel:2003, Anderson:2012}. This is due its ease of implementation and straightforward interpretation; the posterior is given by the number density of chain samples in the long-chain limit.  The principal analysis of temperature data from the Planck satellite used a modified Metropolis-Hastings algorithm, following the methods of \cite{Lewis:2013}, which incorporates a decomposition of the full parameter space into ÔfastÕ and ÔslowÕ subspaces for enhanced speed \citep{PlanckXVI}. 

Metropolis-Hastings sampling has some drawbacks. Choosing the trial distribution relies on our {\it a priori} knowledge of the target distribution and its degeneracies. In many cases this is precisely what we are trying to find through the evaluation of the posterior. Therefore we may be unable to choose an optimal trial distribution, and reaching convergence can take many orders of magnitude longer than indicated by Eq.~\ref{eMHconv}. For a $D$ dimensional parameter space there are $D(D+1)/2$ independent components of the covariance matrix; in high-dimensional parameter spaces a poorly-estimated covariance matrix will result in an impractically long convergence time. Furthermore Metropolis-Hastings suffers from long {\it burn-in} times if one makes an initial choice of parameters which lie far from the bulk of the posterior. Burn-in refers to the initial section of the chain which is not a representative sampling of the posterior. These samples must be discarded in the calculation of the posterior, and thus in some sense require redundant likelihood calls. Moreover, this sampling technique does not naturally return an estimate of the Bayesian evidence.

\subsection{Nested sampling}
\label{ssNS}
Nested sampling was first presented in \cite{Skilling:2004} as a method for Bayesian evidence calculation. This method was developed to sidestep computationally expensive thermodynamic integration (TI) techniques. Nested sampling begins with a sampling of the entire prior volume. Samples are then drawn from successively more likely regions of parameter space until the posterior bulk, which is oversampled with respect to the full prior, is located and explored. The samples are weighted appropriately and the posterior and evidence may be estimated.

Nested sampling was first used in a cosmological application by \cite{Bassett:2004} for discriminating between dark energy models. \cite{Mukherjee:2006} introduced a practical method for sampling from restricted regions of prior space based on using an ellipsoidal approximation to the likelihood contours. This was developed further by \cite{Shaw:2007}, who demonstrated the improved efficiency of nested sampling in evaluating the Bayesian evidence compared to thermodynamic integration methods. \cite{Feroz:2008} and \cite{Feroz:2009} extended single ellipsoidal nested sampling into \texttt{MultiNest}, which deals with multi-modal likelihoods via a cluster-detection algorithm. \cite{Feroz:2013} introduce importance nested sampling which can offer further computational gains. For the sake of performance comparison and analysis we implement our own Python nested sampling routine designed to explore the uni-modal likelihoods encountered in common cosmological models. We outline our practical implementation of the scheme below - see \citealp{Skilling:2004} for detailed discussion of the formalism. 

\subsubsection{Nested sampling formalism and implementation}
We introduce the  {\it prior mass} $X(L)$ associated with likelihoods $\mathcal{L}(\mathbf{\Theta})$ greater than $L$:
\begin{equation}
X(L)=\int_{\mathcal{L}({\bf \Theta})>L}{\pi({\bf \Theta}) d{\bf \Theta}}.
\label{priorvol}
\end{equation}
Clearly $X_0 \equiv X(0) =1$ and $X(L \ge \mathcal{L}_\text{max})=0$. Defining the inverse function $L(X)$ as the likelihood which bounds a prior mass $X$, one can simplify Eq.~\ref{eEvidence} for the Bayesian evidence from a multi-dimensional integral over the prior volume to a one-dimensional integral over the prior mass:
\begin{equation}
\mathcal{Z}=\int_0^1{L(X)dX}.
\label{evidence2}
\end{equation}
Here $dX$ is the prior mass associated with likelihoods in the interval $[L,L+dL]$. The integral is well-defined for continuous $\mathcal{L}$ and $\pi$ with connected support \citep{Chopin:2008}.

Following~\cite{Mukherjee:2006}, our nested sampling scheme works as follows:
\begin{enumerate}[1.]
\item{Draw $N$ points (the {\it active set} $A$) from the prior.}
\item{At iteration $i$ the least likely point in the active set, which we denote ${\mathbf x}_i$ and is such that $\mathcal{L}({\mathbf x}_i)=L_i$, becomes the $i$th sample point.}
\label{NSstep2}
\item{Sample the prior, subject to $\mathcal{L}>L_i$. To do this:}
\begin{enumerate}[-]
\item{Approximate the likelihood contour $\mathcal{L}=L_i$ by an ellipsoid which bounds the active set (see \S \ref{sssDraw}).}
\item{Draw uniformly from the ellipsoid until one has a point ${\mathbf y}$ such that $\mathcal{L}({\mathbf y})>L_i$.}
 \end{enumerate}
\item{This new point ${\mathbf y}$ replaces ${\mathbf x}_i$ in the active set.}
\item{Iterate from step~\ref{NSstep2} until stopping criterion is satisfied.}
\end{enumerate}
\begin{figure}
\centering
\includegraphics[width=85mm]{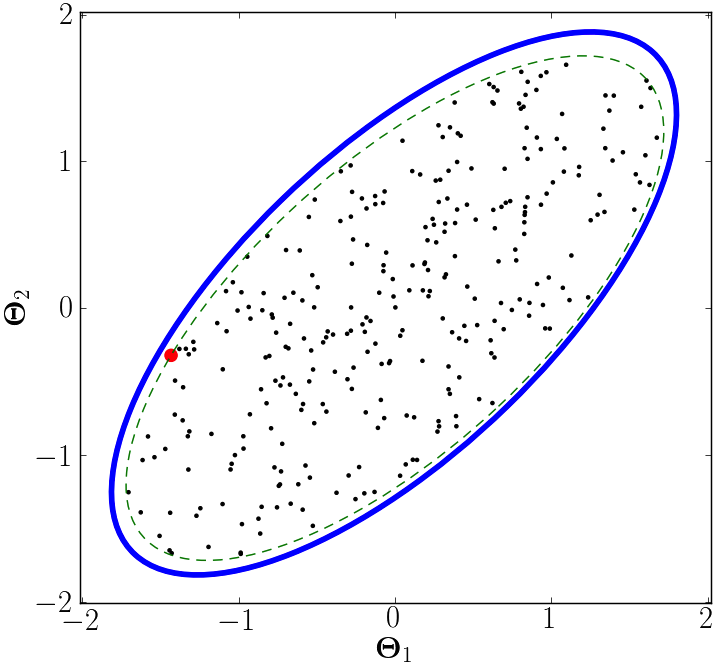}
\caption{{\it Nested sampling}: the active set ($N = 300$, {\it black points}) and the inferred bounding ellipse ({\it blue solid curve}, $f = 1.06$) for a two-parameter Gaussian likelihood ($\sigma_1 = \sigma_2 = 1, \rho = 0.7$). The input mean is $\boldsymbol{\mu}=(0,0)$. The ellipse is inferred from the mean and covariance of the active set (Eq. \ref{eqEllipsoid}). The true likelihood contour ({\it green dashed curve}) corresponding to the lowest likelihood point ({\it red point}) is everywhere encompassed by the ellipse, as necessary for unbiased results.}
\label{figNestSampActSet}
\end{figure}
The initial active set represents $N$ points drawn from the  prior; this is equivalent, by construction of the prior mass $X$, to drawing $N$ values for $X$ uniformly over $[0,1]$. Choosing the lowest likelihood point is equivalent to choosing the largest of these $N$ standard uniform deviates, which has the density
\begin{equation}
p(t)=Nt^{N-1},    
\label{pdf}
\end{equation}
for $t\in[0,1]$; we denote the distribution described by this density as $T$. Thus the prior mass enclosing all points in the active set shrinks to $X_1 = tX_0$, where $t \sim T$. For each subsequent iteration $i$ we choose the lowest likelihood point from which to restrict the prior, and therefore each iteration the prior mass $X_i$ shrinks by a factor $t$, given probabilistically by Eq.~\ref{pdf}.  We obtain a sequence for the remaining prior mass at each iteration:
\begin{equation}
X_0 = 1, X_1 = t_1X_0, X_2 = t_2X_1, ...   
\label{eSequence}
\end{equation}
where the $t_i \sim T$. After many iterations the logarithmic prior mass $\ln (X_i) = \ln(t_1...t_iX_0)$ is the sum of many independent and identically distributed random variables, and so the expectation and standard deviation fully characterise the distribution of $\ln (X_i)$:
\begin{equation}
\ln(X_i)=-\frac{i \pm \sqrt{i}}{N}. 
\label{meanvar}
\end{equation}
Thus the prior mass shrinks exponentially with each iteration; to estimate the evidence and posterior we set
\begin{equation}
X_i=\exp \left(-\frac{i}{N}\right ),
\label{priormassdet}
\end{equation}
and we may straightforwardly propagate the uncertainty on the $X_i$ into an uncertainty on posterior statistics and the final evidence estimate.
The evidence is estimated, by the trapezium rule, as
\begin{equation}
\mathcal{Z}=\sum_{i=1}^{M} L_i w_i + \bar{L}X_M,
\label{evapprox}
\end{equation}
where the $L_i$ are the points of lowest likelihood at each iteration, $w_i=\frac{1}{2}(X_{i+1}-X_{i-1})$ is the prior mass over which $\mathcal{L}(\boldsymbol{\theta}) \approx L_i$ and $M$ is the total number of iterations. The additional term $\bar{L}X_M$ is added to account for the contribution from the active set; we assume each point in the active set occupies an equal fraction of remaining prior volume $X_M$, and $\bar{L}$ is their average likelihood. Including the active set ensures that the peak of the distribution is fully mapped out. The joint posterior distribution may be inferred by binning up the sample points ${\mathbf x}_i$ with weights
\begin{equation}
 \renewcommand{\arraystretch}{2.5}
p_i = \left\{
     \begin{array}{lr}

       \dfrac{L_iw_i}{\mathcal{Z}}, & \hfill i \in \left\{1,...,M \right\},\\
       \dfrac{L_i X_M}{N \mathcal{Z}}, & \hfill i \in \left\{M+1,...,M+N \right\}.
     \end{array}
   \right.
\end{equation}
The  mean and covariance of the posterior can be estimated by
\begin{equation}
\label{eMeanAndCovWeight}
\bar{{\bf x}}=\sum_{i = 1}^{n} p_i \mathbf{x}_i,  \hspace{4  mm}
{\bf C}= \sum_{i = 1}^{n} p_i (\mathbf{x}_i - \bar{{\bf x}})(\mathbf{x}_i - \bar{{\bf x}})^ {\text {\bf T}},
\end{equation}
where $n=M+N$.

\subsubsection{Drawing from a restricted prior}
\label{sssDraw}
At each iteration $i$ we need to sample the prior restricted to the region of parameter space such that $\mathcal{L}(\mathbf{\Theta})>L_i$. Because the prior mass shrinks exponentially with each iteration (Eq.~\ref{priormassdet}), it is important computationally to avoid sampling the whole prior throughout the course of the algorithm, since this would lead to exponentially worsening acceptance rates. \cite{Mukherjee:2006} and \cite{Shaw:2007} suggest using an ellipsoidal approximation to the likelihood contour which is defined by the active set, sampling the prior restricted to this ellipsoid. We summarise this method below, while Fig. \ref{figNestSampActSet} shows the process pictorially.

We define the ellipsoidal approximation to the bounding likelihood contour by 
\begin{equation}
\label{eqEllipsoid}
\mathbf{(x-\boldsymbol\mu)}^{T}\mathbf{C^{-1}(x-\boldsymbol\mu)}=k,
\end{equation}
where $\boldsymbol\mu$ is the centroid, and $\mathbf{C}$ the covariance matrix, of the active set $A$. Also
\begin{equation}
k = \max\left\{(\mathbf{x}_i-\boldsymbol\mu)^{T}\mathbf{C^{-1}}(\mathbf{x}_i-\boldsymbol\mu):\mathbf{x}_i \in A\right\},
\end{equation}
is defined such that the ellipsoid is scaled to encompass the entire active set. We further expand the ellipsoid along each principal axis by a factor $f$ to ensure the entire prior mass $X(L_i)$ is encompassed; this is required for unbiased results. We adopt $f = 1.06$ following~\cite{Shaw:2007}. This expansion is effected by redefining $k \rightarrow kf^2$. Note that correspondingly the volume of the ellipsoid $V \rightarrow Vf^D $.

To sample uniformly from the $D$-dimensional ellipsoid we sample uniformly from the unit $D$-ball and then construct  a linear map $\mathbf{T}$ from the latter to the former. We draw a $D$-dimensional vector of Gaussian random numbers $\mathbf{w}$ then define the unit-vector 
\begin{equation}
\mathbf{z}=\frac{\mathbf{w}}{\mathbf{|w|}},
\end{equation}
which lies on the surface of the unit $D$-ball. A uniform deviate $u \in [0,1]$ then maps $\mathbf{z}$ to a point {\it within} the unit $D$-ball:
\begin{equation}
\mathbf{z} \rightarrow u^{1/D} \mathbf{z}.
\end{equation}
To map this to the ellipsoid the co-ordinates of $\mathbf{z}$ are firstly rotated into the frame of the principal axes of the ellipsoid, then scaled according, and then rotated back. The overall transformation matrix is
\begin{equation}
\mathbf{T} = \sqrt{k}\mathbf{R^TDR},
\end{equation}
where $\mathbf{D}$ is the square root of the diagonalised covariance matrix and $\mathbf{R}$ is the matrix of eigenvectors of $\mathbf{C}$. Finally we shift the origin to lie at the centroid $\boldsymbol\mu$ of the active set. Thus we obtain a deviate $\mathbf{y}$ drawn uniformly from the bounding ellipsoid
\begin{equation}
\mathbf{y} =\mathbf{Tz}+\boldsymbol\mu.
\end{equation}
Nested sampling involves just two tunable hyper-parameters: the number of points in the active set $N$ and the expansion factor $f$. \cite{Shaw:2007} investigated the effect of varying $f$ on the numerical evidence for toy-model cases (in which the evidence can be computed analytically). They found $f=1.06$ sufficient to give unbiased results, and we adopt this as our fiducial value in all cases. As shown in \S \ref{sResults} the size of the active set directly determines the density of sampling in parameter space and, correspondingly, the number of iterations until convergence. Thus $N$ should be chosen large enough so as to provide an accurate approximation to the posterior, whilst being not so large as to invoke impractical computational overheads. Typically $N \sim O(10^2)$ offers this compromise, and we investigate this choice in \S \ref{sResults}.

\subsubsection[sssStop]{Stopping criterion}
\label{sssStop}
\begin{figure}
\centering
\includegraphics[width=85mm]{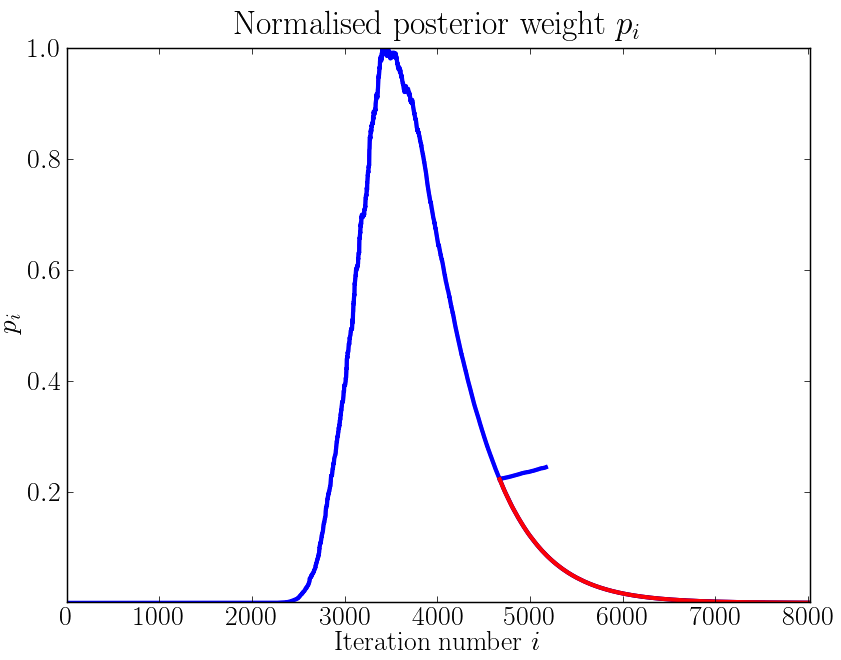}
\caption{{\it Nested sampling}: typical trajectory for posterior weights $p_i$ (or equivalently evidence increments $L_iw_i$) as a function of iteration number $i$ (blue). Note how initially the curve is increasing, as samples are drawn from successively higher likelihood regions. As the active set enters the posterior bulk, around $i \sim 3000$, the likelihood $L_i$ flattens off, the exponentially decreasing prior volume weighting factor $w_i$ becomes dominant and the weights $p_i$ decrease. When the stopping criterion is satisfied (Eq. \ref{stopping}) the points in the active set (sorted arbitrarily from low to high likelihood) are inserted into the chain resulting in the small correction from $i \sim 4700$. The red curve shows the weights trajectory if we do not use the stopping criterion. Here the active set size $N=500$, and truncating the algorithm results in less than a 0.02\% discrepancy on the computed evidence - a negligible effect.}
\label{weights}
\end{figure}

When do we stop sampling the parameter space? For Metropolis-Hastings we use the Gelman-Rubin type statistic $r$, which ensures the variance of the chain mean is much less than the posterior variance for each parameter. Unfortunately this stopping criterion is not appropriate for nested sampling; as a consequence of the sampler traversing the full parameter space, the variance of the sample means can become very small long before we have properly explored the bulk of the posterior, particularly when we use a large active set. Instead we consider the evidence accumulated throughout the course of the algorithm following \cite{Shaw:2007}. The typical trajectory for the values of the posterior weights $p_i$ (or evidence increments $L_iw_i$) can be seen in fig.~\ref{weights}. The peak of this curve corresponds to the exploration of the bulk of the posterior. We use this observation as a guide for when to stop the algorithm: when the points in the active set would, in sum, increment the evidence by only some small fraction of the total evidence accumulated, we can safely assume the posterior bulk has been explored sufficiently. The maximum possible contribution to the evidence from the active set at the $i$th iteration is approximately $\Delta \mathcal{Z}_i=L_{\text{max}}X_i$, and so this condition can be expressed quantitatively as: {\it stop if}
\begin{equation}
\log(\mathcal{Z}_i+\Delta \mathcal{Z}_i)-\log(\mathcal{Z}_i)<\kappa. 
\label{stopping}
\end{equation}
Here $L_{\text{max}}$ is the maximum likelihood point in $A$ and we take $\kappa = 0.1$, which is small enough that the error on the accumulated evidence due to truncation is negligible compared to the uncertainty resulting from the deterministic approximation for $X_i$ (Eq.~\ref{priormassdet}) - see~\cite{Chopin:2008}. 

\subsection{Affine-invariant ensemble MCMC}
\label{ssAffInv}
\cite{Goodman:2010} introduced an ensemble (many-particle) MCMC sampling algorithm which has the property of affine-invariance; that is, the performance of the algorithm is invariant under linear transformations of the parameter space. Thus, in particular, the sampler works just as well on a highly degenerate Gaussian distribution as an uncorrelated and isotropic Gaussian distribution. The principle is that many particles, or {\it walkers}, move through parameter space; at each iteration each walker undergoes a trial move, with the step being accepted with some probability (described below). The trial move is based on the positions of each of the other walkers (the complementary set), since these provide information about the underlying distribution. 

\begin{figure}
\centering
\includegraphics[width=85mm]{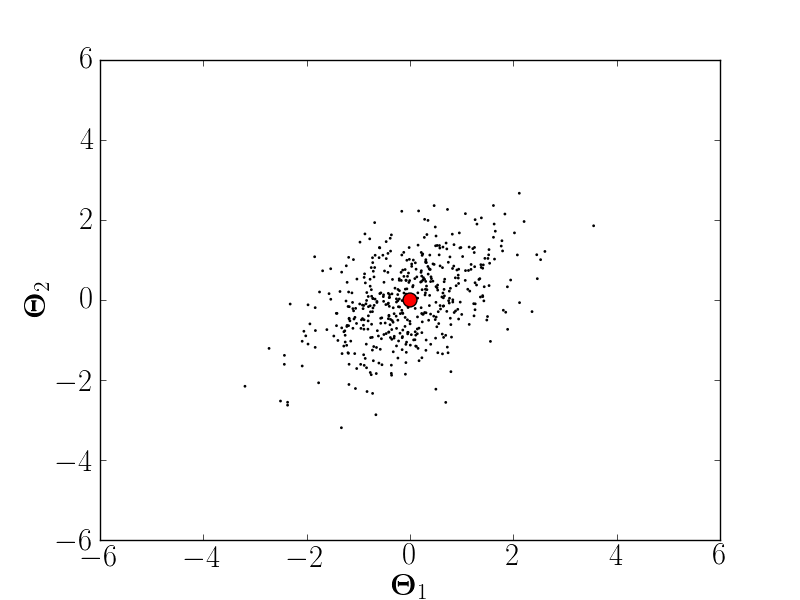}
\caption{{\it Affine-invariant MCMC}: snapshot of the positions of the $n = 500$ walkers (black points) after $t=100$ steps (per walker) for a two-parameter Gaussian likelihood ($\sigma_1 = \sigma_2 = 1, \rho = 0.4$). The input mean $\boldsymbol{\mu}=(0,0)$ has been highlighted (red dot).}
\label{figAffInv}
\end{figure}

We describe the algorithmic scheme below for a one-dimensional density $p(x)$ (see also \citealp{Foreman-Mackey:2013}):
\begin{enumerate}[1.]
\item{The positions of the $n$ walkers are initialised ($t=0$). Suppose at iteration $t$ the positions of all the walkers are described by ${\mathbf x}(t)$}
\item{For each of the walkers $x_j(t), j = 1,..., n$ successively:}
\label{step2affinv}
\begin{enumerate}[-]
\item{Propose a trial step $x_\text{trial}$ (the {\it stretch move}): 
\begin{equation}
x_\text{trial} = x_k + z(x_j(t) - x_k),
\end{equation}
where $x_k$ is a random walker from ${\mathbf x}_{[j]}(t)$, the set of positions of the other walkers, and we draw $z$ from
\begin{equation}
g(z) \propto z^{-\frac{1}{2}}, \hspace{2mm}z \in \left[ \frac{1}{a}, a \right].
\end{equation}}
\item{Define $s = \min \left\{1, z^{D-1}p(x_\text{trial})/p(x_j(t)) \right\}$. Assign
\begin{equation}
x_j(t+1) \leftarrow \left\{
     \begin{array}{lr}
       x_\text{trial} & \hfill \text{with prob. }  s\\
       x_j(t) & \hfill \text{with prob. } 1-s
     \end{array}
   \right.
\end{equation}
}
\end{enumerate}
\item{Iterate over $t$ from step~\ref{step2affinv} to obtain $\left\{{\mathbf x}(t)\right\}$.}
\end{enumerate}
Note that we set the step-size parameter $a = 2$ in all cases, but in principle $a$ may be varied if the acceptance fraction is too low or too high (see ~\cite{Goodman:2010} and ~\cite{Foreman-Mackey:2013} for discussion). The above form of $g(z)$ makes the trial step symmetric and, in conjunction with the modified Metropolis-Hasting acceptance probability $s$, ensures detailed balance and therefore that the asymptotic distribution is $p$. See Fig. \ref{figAffInv} for an example of walker positions at one particular time-step long after burn-in for a degenerate Gaussian likelihood.

For affine-invariant MCMC we use the autocorrelation time to define the stopping criterion, following \cite{Goodman:2010} and ~\cite{Foreman-Mackey:2013}. One must be careful to distinguish burn-in (where the mean walker position can vary considerably) from when the walkers are properly sampling the posterior (where the mean position makes only small oscillations around the mean parameters). The autocorrelation time measures the longest wavelength of these oscillations, and corresponds to the number of steps between independent samples of the posterior. Burn-in is estimated from the {\it exponential} autocorrelation time $\tau_{\rm exp}$ for the mean position of the walkers at each subsequent iteration, found by a least squares fit to the autocorrelation function. We cut $b = 5 \tau_{\rm exp}$ steps as burn-in to ensure the initialisation is forgotten, and then estimate in the same way the autocorrelation time $\tau$ for the post burn-in samples from the remaining steps (see \S \ref{ssComp} for more details). The {\it integrated} autocorrelation time $\tau_{\rm int}$  measures the variance on the mean for each parameter $x$; if after burn-in we have $N$ samples in total:
\begin{equation}
\label{eqAutoCorr}
\sigma^2_{\bar x} = \frac{2\tau_{\rm int}}{N}\sigma^2_0.
\end{equation}
We find, as in~\cite{Akeret:2012}, that the autocorrelation function has an exponential form and so we may set $\tau = \tau_{\rm exp} = \tau_{\rm int}$. Thus, using Eq.~\ref{er} in Eq.~\ref{eqAutoCorr}, the total number of likelihood evaluations for convergence is
\begin{equation}
N_{\rm like} =\frac{2 \tau}{r}+bn.
\label{eNlikeAffInv}
\end{equation}
As for Metropolis-Hastings we demand $r = 0.01$ for convergence, so we may make a direct comparison between the two methods. 

The mean and covariance of a chain of walkers are given as for Metropolis-Hasting by Eq.~\ref{eMeanAndCov}, where only the samples after $b$ steps are used.
Affine-invariant ensemble MCMC uses the positions of the walkers at each step to provide information about where to sample next. A curving (`banana-shape') distribution would result in a low efficiency for Metropolis-Hastings because the trial distribution cannot be tuned throughout the parameter space (the optimal trial covariance varies as function of position). However for this multiple-walker technique, the positions of the walkers ensures trial steps throughout parameter space are restricted to the posterior bulk, and therefore that the acceptance probability is sufficiently high for practical applications. The only tunable parameters for this sampling technique are the step-size parameter $a$ and the number of walkers $n$. The number of walkers can be chosen to suit the application, but typically $n \sim O(10)$ or $O(10^2)$. 

\section{Toy-model (Gaussian) likelihoods}
\label{sResults}
In this section we focus on the number of likelihood evaluations required for convergence, $N_{\rm like}$, for each of the three sampling methods, for the case of simple Gaussian likelihoods. In models where likelihood calls are expensive, $N_{\rm like}$ is a diagnostic quantity for the computational effort required for each of these sampling methods. Example Python code - for nested and affine-invariant sampling - has been made publicly available; this allows the user to explore the toy-model Gaussian likelihoods studied this section\footnote{https://github.com/rupert-allison/sampling}.

\subsection{An analytical result for nested sampling}
We firstly derive a new analytical estimate for the expected number of likelihood evaluations required for convergence in nested sampling, in the case of a multivariate Gaussian likelihood and uniform priors. The result, although strictly holding only for this special case, is - as seen below - nevertheless useful for order of magnitude predictions for more complicated distributions. We begin with the stopping criterion in Eq.~\ref{stopping}. Let the sampler converge after $M$ iterations; then $M$ satisfies $\Delta \mathcal{Z}_M/\mathcal{Z}_M = s$ for constant $s=e^\kappa - 1 \approx 0.1$. Using $\Delta \mathcal{Z}_M=L_{\text{max}}X_M$ and $\mathcal{Z}_M \approx \mathcal{Z}$, the total number of iterations $M$ will approximately satisfy
\begin{equation}
M=N\ln\left(\frac{L_{\text{max}}}{s \mathcal{Z}}\right).
\label{Miter}
\end{equation}
Consider a Gaussian likelihood 
\begin{equation}
\mathcal{L}({\mathbf x})=\frac{1}{\sqrt{(2\pi)^D \det {\mathbf C}}}\exp \left[-(\mathbf{x}-\mbox{\boldmath$\mu$})^T{\mathbf{C}^{-1}}(\mathbf{x}-\mbox{\boldmath$\mu$})\right],
\end{equation}
and a uniform prior of volume $V_\text{p}$ which comfortably contains the bulk of the posterior, so that $\mathcal{Z} = 1/V_\text{p}$. Then near convergence $L_{\text{max}} \approx \mathcal{L}(\mbox{\boldmath$\mu$})$, so Eq.~\ref{Miter} becomes
\begin{equation}
M = N \ln \left( \frac{V_\text{p}}{V_\text{t}s } \right),
\label{Miter2}
\end{equation} 
where $V_\text{t} = \sqrt{(2\pi)^D \det \boldsymbol{C}}$ is a measure of the volume of the target (posterior) distribution. 

At each iteration $i$ the ellipsoidal bound will not necessarily be congruous with the likelihood contour $\mathcal{L}=L_i$; indeed, for unbiased results we require only that the locus of the ellipsoid satisfies $\mathcal{L} \le L_i$ everywhere. Thus some samples, drawn uniformly from the interior of the ellipsoid, will also be drawn from $\mathcal{L}<L_i$, and these must be rejected. In order to calculate the number of likelihood evaluations required for convergence we need to understand this acceptance rate. In the case of uniform priors, the acceptance rate is essentially the ratio of the volume of the restricted prior ($\mathcal{L}>L_i$) which we denote by $V_\text{like}$, to the volume of the bounding ellipsoid $V_\text{ell}$. Since we always scale the ellipsoid until it encompasses the active set, and because we expand each principal axis by a factor $f$, we parameterise the volume ratio - and thus the acceptance rate AR - by
\begin{equation}
\label{AR}
\text{AR} = \frac{V_{\text{like}}}{V_{\text{ell}}} = \left( \frac{\alpha}{f} \right)^D.
\end{equation}
Here $\alpha$ represents the mean ratio between the linear dimension of the likelihood contour and the ellipsoid before we expand by $f$. We expect this parameter to increase with increasing $N$ and decreasing $D$ as the constraints on the covariance improve. Using this parameterisation and Eq.~\ref{Miter} we can finally write down an expression for the number of likelihood evaluations until convergence, $N_\text{like}$:
\begin{equation}
\label{Nlike}
N_{\text{like}} = N \left( \left( \frac{f}{\alpha} \right)^D \ln \left( \frac{V_\text{p}}{V_\text{t}s } \right)+1 \right),
\end{equation}
since $N_\text{like}=M/\text{AR}+N$, where the last term accounts for the active set. 

To estimate how $\alpha$ depends on $N$ and $D$, we performed simple numerical studies based on drawing $N$ points uniformly from an ellipsoid then comparing the volume of the inferred bounding ellipsoid to the actual volume bounded by the lowest likelihood contour. We find that for Gaussian likelihoods, in general, $\alpha \in [0.92,1]$. The dependance on $N$ and $D$ is weak, justifying the parameterisation used in Eq. \ref{AR}. Setting $\alpha = 0.92$ in Eq.~\ref{Nlike} gives an upper bound on $N_{\rm like}$ for Gaussian likelihoods.

Since in general $V_{\rm p}/V_{\rm t}$ scales exponentially with dimension $D$, the logarithmic term gives a contribution linear in $D$, which is the same scaling as Metropolis-Hastings. The additional exponential term deriving from the imperfect estimation of the covariance from the active set means that convergence for nested sampling scales more severely with dimension.

\begin{figure}
\centering
\includegraphics[width=85mm]{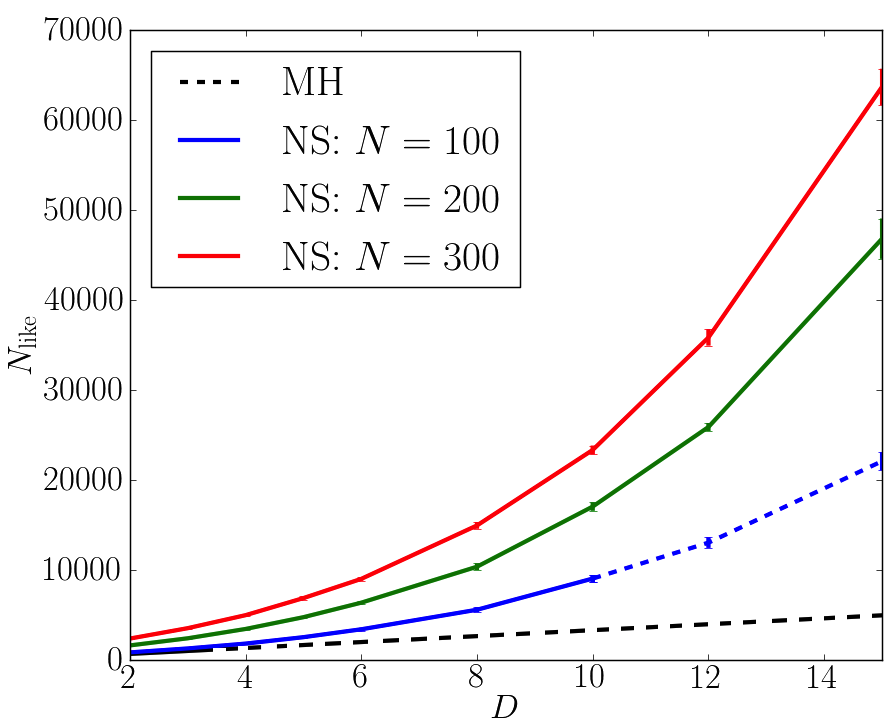}
\caption{Number of likelihood evaluations for convergence, $N_{\rm like}$, as a function of parameter space dimension $D$ for optimal Metropolis-Hastings (MH) and nested sampling (NS). Here we consider an isotropic Gaussian likelihood and a conservative (large) value for the prior to posterior volume ratio of $(20/\sqrt{2 \pi})^D$. The error bars for the NS curves show the standard deviation from multiple runs. For MH we show the result  of Dunkley et al. (2005): $N_{\rm like} = 330D$, which ensures $r<0.01$. For NS the time for convergence is a linear function of the size of the active set $N$, and the steepening gradient with dimension is because probability mass is more concentrated at larger radii in higher dimensions. These curves can be modelled by Eq.~\ref{Nlike}. We dash the $N=100$ curve for high dimensions because the posterior statistics are likely unreliable in this regime where the active set does not accurately constrain the covariance.}
\label{figNS+MH}
\end{figure}

\subsection{Comparison of sampling methods on Gaussian likelihoods}
Here we implement and compare each of the three sampling techniques on Gaussian toy-model likelihoods. We consider Metropolis-Hastings sampling where the trial covariance matrix is optimally tuned (Eq. \ref{tuning}). Poorly-tuned schemes can be computationally prohibitive even in the isotropic Gaussian regime, e.g., a factor of 2 under-estimation in the trial step-size in each of 7 dimensions would require a factor of over $10^2$ more iterations for convergence than in the optimally-tuned case; see e.g. Fig. 9 of \cite{Dunkley:2005}. Additionally, not properly accounting for parameter degeneracies will further lengthen the convergence time.

Fig.~\ref{figNS+MH} demonstrates the non-linear scaling of nested sampling with the dimensionality of the parameter space. We find that the form of these curves is accurately modelled by the analytic result given in Eq. \ref{Nlike}. This effect arises  due to the concentration of probability mass at larger radii in higher dimensional spaces, and so more samples are drawn from the region (within the ellipsoid) where ${\mathcal L}<L_i$, reducing the acceptance rate. As seen in Eq.~\ref{Nlike} the convergence time only depends logarithmically on the prior to posterior volume ratio, and thus should be insensitive to the particular choice of the extent of the prior. Larger active sets take longer to converge, but provide correspondingly more posterior samples, and hence less statistical uncertainty on the posterior statistics. More quantitatively, the posterior bulk is reached in $N_{\rm tar} \sim N \ln(V_{\rm p}/V_{\rm t})$ iterations, so the effective number of samples of the posterior is $N_{\rm eff} = M-N_{\rm tar}+N \sim N \ln(e/s)$. Thus larger active sets should be chosen for a denser sampling of the posterior (and hence less noisy estimates of the mean and covariance). For quick parameter constraints the smallest possible active set should be chosen, subject to $N \gg D$. This condition ensures that the ellipsoidal bound contains the entire restricted prior volume, and thus provides unbiased results.

\begin{figure}
\centering
\includegraphics[width=85mm]{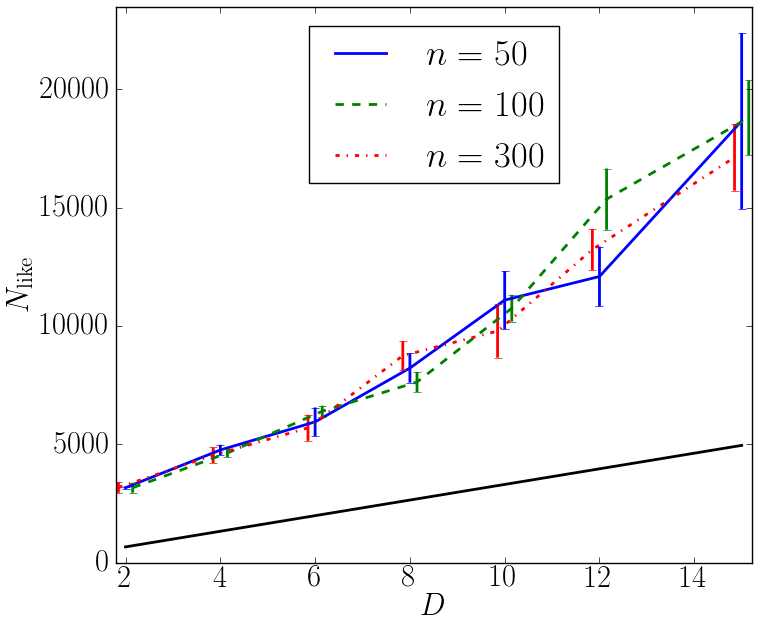}
\caption{Number of likelihood evaluations for convergence $N_{\rm like}$ as a function of parameter space dimension $D$, for optimal Metropolis-Hastings (MH) and affine-invariant MCMC (Aff-Inv). Here we consider a toy-model isotropic Gaussian likelihood and assume negligible burn-in time. Note that Aff-Inv requires more likelihood evaluations than optimal MH for convergence to the same level of accuracy on the parameter means ($r = 0.01$); this is due to a longer autocorrelation time for the chain. However, for a poorly tuned trial distribution the convergence time can be an order of magnitude longer for MH than Aff-Inv, which requires essentially no tuning. The convergence time for Aff-Inv is independent of the number of walkers $n$ since we neglect burn-in. We have slightly displaced the $n=100$ and $n=300$ data points from integer dimension $D$ for visual clarity. The error bars represent the 1$\sigma$ sampling uncertainty derived from multiple runs.}
\label{figAffInv+MH}
\end{figure}
Fig.~\ref{figAffInv+MH} shows the convergence time $N_{\rm like}$ (to achieve $r = 0.01$) for optimal Metropolis-Hastings and affine-invariant MCMC on a toy-model isotropic Gaussian likelihood. The number of likelihood evaluations required is higher for affine-invariant MCMC than for Metropolis-Hastings. Non-optimal Metropolis-Hastings, however, can give significantly longer convergence times. Because we neglect burn-in by starting the walkers from a sampling of the posterior, $N_{\rm like}$ is independent of the number of walkers $n$. The convergence time for affine-invariant MCMC is lower than nested sampling for all high dimensional spaces, but nested sampling can give less statistical noise on the posterior statistics for the same computational effort. 

\section[cosmoparams]{Cosmological parameters}
\label{sCosmoParams}
Here we test the performance of each of the three sampling techniques on real data. We focus on several performance diagnostics: the number of likelihood evaluations required for convergence, the scope for parallelisation, dimensional scaling, requisite tunings and performance on non-Gaussian distributions. 

\subsection{The data and the model}
\label{ssData}
We consider Cosmic Microwave Background (CMB) data from the Wilkinson Microwave Anisotropy Probe (WMAP) 7 year observations \citep{Jarosik:2011}. The data we use are measurements of the maps and angular power spectra of the CMB anisotropies in temperature and polarisation \citep{Larson:2011}. We adapt the popular \texttt{CosmoMC} code, replacing the default Metropolis-Hastings sampler by Python implementations of nested sampling and the Goodman \& Weare affine-invariant ensemble sampler. The Python samplers are called within the Fortran \texttt{CosmoMC}, giving us the usability of Python (for the modified code) whilst retaining the speed and robustness of the well-tested Fortran code in the computationally intensive part of the calculation. 

The chosen sampler passes a set of cosmological parameters to \texttt{CAMB} \citep{Lewis:1999}: this code then integrates the relevant Boltzmann equations  - which encode the physics of the production and propagation of the CMB - and computes theory power spectra, given this parameter set. The WMAP likelihood code then computes how well the theory explain the data, and this number is passed back to the sampler which then chooses a new parameter set (as explained in \S \ref{sSampling}). This process continues until the stopping criterion is satisfied. 

For comparison purposes we consider the concordance $\Lambda$CDM model, which is completely described by 6 parameters. We take as a basis $\left\{ \Omega_b h^2, \Omega_c h^2, \theta, \tau, n_s, \ln (10^{10}A_s) \right\}$: the physical baryonic density, the physical cold dark matter density, the angular size of the sound horizon at recombination, the optical depth to reionisation, the scalar spectral index and the logarithmic amplitude of the initial scalar fluctuations (at pivot scale $k_0=0.002$ Mpc$^{-1}$), respectively. This choice is  convenient since the posterior is approximately Gaussian for each parameter. We also include an additional parameter $A_{\rm SZ}$ describing the amplitude of the Sunyaev-Zel'dovich effect, which increases the observed power in CMB fluctuations across a wide range of angular scales \citep{Larson:2011}.  The most recent all-sky CMB data now come from the Planck satellite \citep{PlanckXV}. The resolution and sensitivity of the data available from Planck mean many more foreground and nuisance parameters may be included in the generative models. The increased number of free parameters makes nested sampling less competitive in this case; see \S\ref{ssParallel} for further discussion. 

\subsection{Comparison of the sampling methods}
\label{ssComp}
\begin{table*}
\begin{center}
\begin{tabular}{|r|r|r|r|r|r|r|r|r|r|r|r|r|}
\hline \hline
$$ & $\hphantom{c}$ & \multicolumn{2}{c|}{MH} & $\hphantom{c}$ & \multicolumn{4}{c}{NS} & $\hphantom{c}$ & \multicolumn{3}{|c|}{Aff-Inv} \\ \hline
$$ & $$ & Optimal & Non-optimal & $$ & $N=50$ & $N=100$ & $N=200$ & $N=300$ & $$ & $n=30$ & $n=50$ & $n=100$  \\ 
$N_{\rm like}$ & $$ & $6400$ & $110800$ & $$ & $2300$ & $5400$ & $11600$ & $17900$ & $$ & $17900$ & $22900$ & $35800$  \\ 
$M$ & $$ & $6400$ & $110800$ & $$ & $900$ & $1800$ & $3600$ & $5300$ & $$ & $17900$ & $22900$ & $35800$  \\ 
AR & $$ & $0.35$ & $0.09$ & $$ & $0.38$ & $0.33$ & $0.31$ & $0.30$ & $$ & $0.48$ & $0.45$ & $0.46$  \\ 
$r$ & $$ & $0.0095$ & $0.0099$ &$$ & $0.0053$ & $0.0023$ & $0.0009$ & $0.0008$ & $$ & $0.01$ & $0.01$ & $0.01$  \\ 
\hline
\end{tabular}
\end{center}
\caption{Number of likelihood evaluations $N_{\rm like}$ for convergence for each of Metropolis-Hastings (MH), nested sampling (NS) and affine-invariant sampling (Aff-Inv), for a 7-parameter $\Lambda$CDM model with WMAP 7 year CMB data. We list the acceptance rate AR, total number of samples $M$ and the normalised measure $r$ of the variance on the parameter means (Eq. \ref{er}). `Optimal' MH uses a well-tuned trial distribution, non-optimal MH uses a trial distribution with optimal parameter widths but no correlation between parameters, resulting in a much longer convergence time (see text \S \ref{ssComp}). NS with $N=50,100$ produces more accurate posterior statistics (as measured by $r$) in a shorter convergence time $N_{\rm like}$ than MH. Aff-Inv sampling requires the most likelihood evaluations due to a long burn-in period.}
\label{tabNlike}
\end{table*}

\begin{figure*}
\centering
\makebox[\textwidth][c]{\includegraphics[width=190mm]{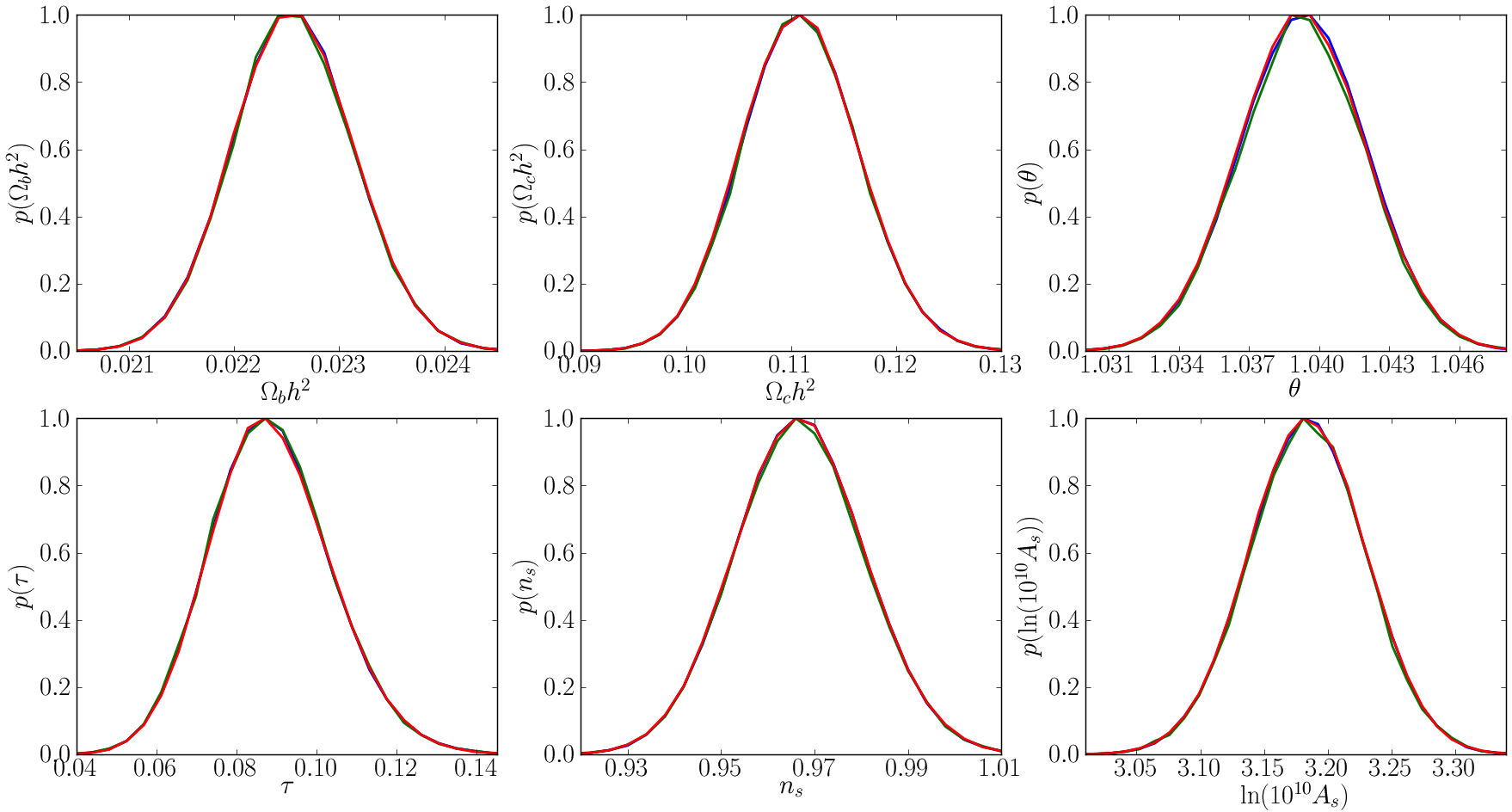}}
\makebox[\textwidth][c]{\includegraphics[width=190mm]{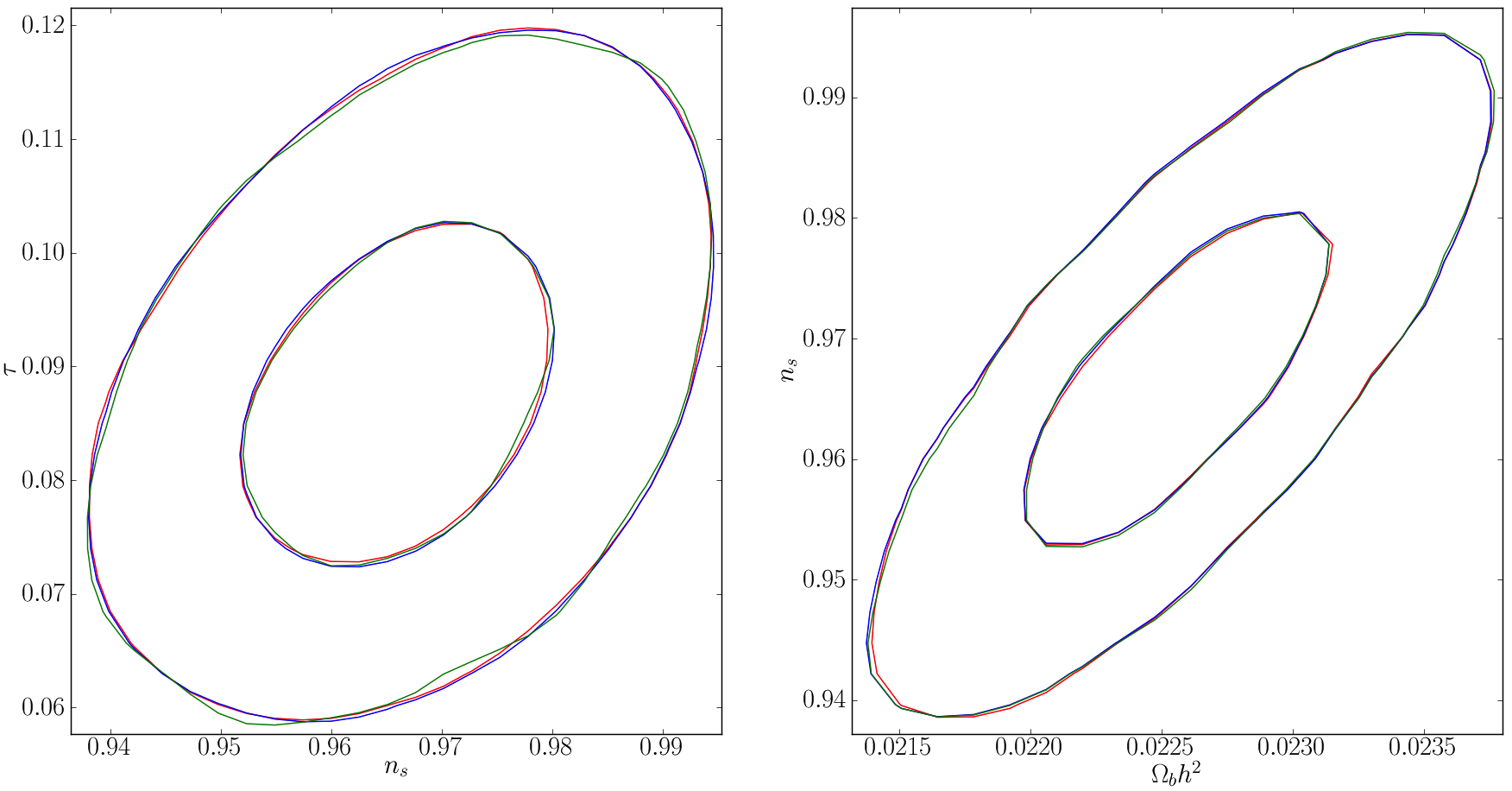}}
\caption{{\it Top panel:} Marginalised posterior probability distributions for each of the basis cosmological parameters (excluding $A_{\rm SZ}$). {\it Bottom panel:} Examples of posterior contours (68\% and 95\% CL shown) demonstrating parameter degeneracies. We show the distributions for each of the three sampling methods: Metropolis-Hastings ({\it blue}), nested sampling ($N=500$) ({\it green}) and affine-invariant MCMC ($n=50$) ({\it red}). There is excellent agreement between the three approaches. The plots are produced from chains ran past the time of convergence; for each method noisier posterior plots result from using only samples up to the convergence time.}
\label{figComparison}
\end{figure*}

We plot the marginalised one-dimensional posterior probability distributions for each of the basis parameters in fig.~\ref{figComparison}. We also show examples of two-dimensional posterior contours demonstrating parameter degeneracies seen in the data. All three methods show excellent mutual agreement and the derived means and constraints are fully consistent with those published by the WMAP collaboration\footnote{http://lambda.gsfc.nasa.gov/product/map/dr4/parameters.cfm}. We do not show $A_{\rm SZ}$, which is uniform across the entire prior range, because the WMAP data do not constrain this parameter. 

Table~\ref{tabNlike} lists the convergence time $N_{\rm like}$ for each of the the three sampling routines based on the  WMAP likelihood. The trial covariance matrix for Metropolis-Hastings was obtained as the output covariance matrix from a previous analysis of this data, and thus is close to optimal. We neglect the number of likelihood evaluations to obtain the trial distribution in the quoted result, and as such this number should be regarded as a lower bound. The initial sample point for Metropolis-Hastings was drawn from a non-degenerate Gaussian with widths close to the posterior width for each parameter. For affine-invariant MCMC we chose random positions for the walkers in a small 7-dimensional ball around the WMAP 7 year best-fit parameters.  

We quote the number of likelihood evaluations for convergence in a single chain. However, to derive estimates of $r$ for Metropolis-Hastings and nested sampling, we actually ran several chains. For affine-invariant MCMC we set $r=0.01$ to determine $N_{\rm like}$; see Eq. \ref{eNlikeAffInv}. 

Nested sampling on the WMAP data performs comparably to the toy-model case in $7$ dimensions, since the posterior is approximately Gaussian for most parameters and the sampler does not require a particular tuning. By comparing the input trial covariance matrix for Metropolis-Hastings sampling with the posterior covariance we find the trial distribution is close to optimal; the fractional error on the input marginalised parameter widths is less than $2$\% for all parameters, and the absolute errors on the input trial correlation coefficients are less than 0.04. Thus the input covariance can be said to be nearly optimal. As such, we infer that the non-Gaussianity of the posterior (particularly for $A_{\rm SZ}$, which has a uniform distribution) is the principal reason for the factor of three increase in convergence time from the optimal toy-model case. 

We ran Metropolis-Hastings using a Gaussian trial distribution with optimal widths for each parameter but assuming no degeneracy between parameters; the number of likelihood evaluations for convergence was a factor of $\sim 20$ greater than in the optimal case, demonstrating the sensitivity of Metropolis-Hastings to the choice of trial distribution (see Table 1).

The nested sampling runs deliver much lower numerical uncertainty on the means, as measured by $r$, compared to both Metropolis-Hastings and affine-invariant MCMC. Small active sets (e.g. $N = 50$) estimate the mean well, yet the tails of the distribution are poorly estimated if $N$ is chosen to be too small, which leads to very noisy estimates of the covariance and the evidence. The time for convergence on real data increases with the number of walkers for affine-invariant MCMC. This is because the burn-in becomes increasingly significant; by $n=100$ burn-in is the dominant phase of the algorithm. The increase in the convergence time, with respect to the toy-model case in $7$ dimensions (fig.~\ref{figAffInv+MH}), is  consistent with the inclusion of burn-in i.e., the autocorrelation times for the sampler are comparable between the two cases. 

To determine the autocorrelation time $\tau$ for affine-invariant sampling we make a least-squares fit to the autocorrelation function which is estimated using the Python script \texttt{acor} \footnote{https://github.com/dfm/acor}.  By comparison with estimates made using long chains of $O(10^4)$ steps, we find estimates for $\tau$ are order of magnitude correct within a few thousand steps (the burn-in samples must be removed for an accurate estimate). The estimated autocorrelation time is then put into Eq.~\ref{eNlikeAffInv} to give the convergence times $N_{\rm like}$ as given in Table~\ref{tabNlike}. 

Despite the large number of likelihood calls required for convergence with affine-invariant MCMC, the walker steps can be made in parallel and so if one has access to a computing cluster the effective time for a likelihood call can be lower by up to a factor $n/2$  (see discussion in \S \ref{ssParallel}). Thus if $n$ is large, although burn-in is longer, affine-invariant MCMC may in practice perform parameter estimation most quickly out of the three samplers. 

\section{Discussion}
\label{sDiscussion}
We discuss the performance of the three sampling techniques on toy-model and cosmological data. We discuss the tunings, convergence time, scope for parallelisation and robustness of each technique for different parameter estimation problems.

\subsection{Tunings and convergence}
The Metropolis-Hasting algorithm convergence time depends strongly upon the choice of trial distribution. For the results derived in \S \ref{sCosmoParams} the trial covariance matrix was derived from the output of a previously converged chain for the same model, and so was close to optimal. A trial distribution which is not tuned to the particular problem - for example an isotropic Gaussian trial used on a highly degenerate target distribution - will fail to return a good sampling of the posterior within a practicable time-scale. For a Gaussian likelihood in a $D$ dimensional parameter space we must specify the $D(D+1)/2$ independent elements of the target covariance matrix to high accuracy to ensure an optimal convergence rate. This requires some knowledge of the underlying posterior distribution, which we may not have access to {\it a priori}. In practice we might run and analyse several short chains, in an inevitably subjective manner, to provide a rough estimate of the covariance before the full problem is tackled. Another option is to use an adaptive Metropolis-Hastings procedure which allows the trial covariance to be learned during the initial phase of the exploration \cite[e.g.,][]{Lewis:2013}. Non-Gaussian likelihoods clearly present an increased level of difficulty for the Metropolis-Hastings algorithm, and widely spread modes of a posterior may fail to be found at all. However, optimal Metropolis-Hastings on uni-modal Gaussian posteriors of a known shape, due to the simplicity and ease of implementation of the algorithm, is a plausible option when massive parallelisation is unfeasible or unnecessary. 

Nested sampling requires no initial guess for the posterior covariance. Instead the algorithm firstly probes the full prior volume, followed by successively more likely regions of parameter space, adapting - at each step - to the inferred shape of the posterior. All samples taken outside the posterior, i.e. the first $N_{\rm tar} \sim N \ln(V_{\rm p}/V_{\rm t})$ iterations, are effectively burn-in since they contribute negligible weight to the posterior. The only tunable hyper-parameters are the size of the active set $N$ and the factor $f$ by which the ellipsoidal bound is expanded to ensure unbiased results. We find $f=1.06$ to be sufficient for all the likelihoods we have considered, while $N$ may tailored to suit one's application. A small active set will provide quick (but noisy) parameter constraints, but a large active set densely samples the posterior; Eq.~\ref{Nlike} allows us to estimate the computational cost for either case. Because of the lack of tuning needed, nested sampling is in some sense more objective than Metropolis-Hastings, and perhaps more straightforward to set-up on a new problem.

Approximating the likelihood contour as a single ellipsoid, as presented here, would give poor performance on multi-modal posteriors since the acceptance rate would drop dramatically once the ellipsoid encountered multiple peaks. Cluster detection algorithms have been incorporated into nested sampling implementations to resolve this issue \citep{Feroz:2009, Feroz:2013}. These extensions also allow nested sampling to be used on distributions containing curving degeneracies, but come with added cost of a more complex implementation and interpretation.

Nested sampling performs well on the cosmological data set, producing high fidelity estimates of the posterior statistics in a small number of likelihood calls. Nested sampling is the only sampler presented here that returns a robust estimate for the evidence. This is because MCMC methods do not explore the entire prior volume in finite time, and the tails of the likelihood may give a large contribution to the evidence integrand. Thus nested sampling is an invaluable tool for model selection analyses \citep[e.g.,][]{Mukherjee:2008}. 

Affine-invariant ensemble MCMC also comes with just two tunable hyper-parameters: the number of walkers $n$ and the step-size parameter $a$, which we set equal to 2 (\S \ref{ssAffInv}). For any given walker $W$ the positions of the other walkers provides information on the direction in which $W$ should move to stay within the posterior bulk. A large number of walkers results in a long burn-in time, but once this period is over just a few steps (for each walker) is enough for robust parameter estimation, since after burn-in each walker is an independent sample from the posterior. As pointed out in~\cite{Foreman-Mackey:2013}, choosing a large number of walkers increases the scope for parallelisation (\S \ref{ssParallel}).

This sampling prescription, by construction, performs equally well on distributions with strong parameter degeneracies as on isotropic distributions, requiring no extra tuning from the user. Provided the initial walker positions are sufficiently dispersed throughout the prior volume, multi-modal and curving distributions are naturally probed by multi-particle samplers such as affine-invariant ensemble MCMC. Affine-invariant MCMC works well on the toy-model Gaussian likelihoods but requires many more likelihood calls than nested sampling on the real data set. Despite this, parallelisation of the walker moves and its potential for mapping out multi-modal or curving distributions ensures this is a competitive sampling method. We point out in~\ref{ssComp}, however, that there are limitations on using the autocorrelation time either as a determination of the end of burn-in or to define a stopping criterion. 

\subsection{Parallelisation}
\label{ssParallel}
Metropolis-Hastings sampling can be parallelised by running multiple, independent chains on separate processors. If a single processor requires $N_{\rm like}$ steps for convergence, then $n_{\rm chain}$ processors can achieve the same accuracy $r$ on the posterior means with $N_{\rm like}/n_{\rm chain}$ samples per processor, reducing the actual convergence time significantly. Alternatively, if all processors are ran for $N_{\rm like}$ steps, one obtains an accuracy $r/n_{\rm chain}$ on the posterior mean since there are a factor of $n_{\rm chain}$ more samples, which produces less noisy parameter constraints and smoother distributions.

The number of likelihood evaluations for convergence in nested sampling is set principally by the size of the active set $N$. One cannot run multiple chains to reduce the number of likelihood evaluations per processor as in Metropolis-Hastings, since the active set is not independent from step to step. However, multiple processors can be used to produce dense samplings of the posterior by combining the samples from multiple chains with appropriate weights. As pointed out in Feroz, Hobson \& Bridges (2009) and implemented in \texttt{MultiNest}, an acceptance rate of less than unity means that parallel sampling from the ellipsoid at each iteration can significantly reduce the wall time of nested sampling. This allows it to be used in high-dimensional parameter spaces where otherwise the exponential scaling (Eq. \ref{Nlike}) would rule it out as a practical sampling method. This is utilised in \cite{PlanckXXII}, where the full generative model includes more than a dozen foreground and nuisance parameters, to explore spaces with $D>20$ parameters.

\cite{Foreman-Mackey:2013} implement parallelisation of the affine-invariant walker moves in their Python code \texttt{emcee}. At each step the walkers can be partitioned into two sets containing $n/2$ walkers each; one set is then held fixed while the positions of the other set of walkers are updated, moving them according only to the positions of the fixed set of walkers. Each of the $n/2$ walker moves can be made in parallel because they are independent. The roles of the two sets are then reversed to complete one step. This provides scope for performing very fast parameter estimation, utilising up to $n/2$ processors. This parallelisation is investigated on cosmological data in~\cite{Akeret:2012}. The dependence of trial moves on the other walkers' positions (c.f. independence of Metropolis-Hastings chains) is precisely the property which allows affine-invariant MCMC to be successful on non-Gaussian and curving distributions. 

\section{Conclusions}
\label{sConc}
We have outlined three sampling methods, presenting the algorithms themselves, discussing their practical implementation and associated stopping criteria. We present a new analytical result for the number of likelihood evaluations required for convergence in nested sampling.  A comparison is made between the three sampling methods, focusing on performance on toy-model Gaussian likelihoods and a dataset derived from measurements of the Cosmic Microwave Background. We make use of and adapt the widely-used cosmological parameter estimation code \texttt{CosmoMC} \citep{Lewis:2002}, developing a general C/Cython bridge between the Fortran \texttt{CosmoMC} and Python implementations of the sampling algorithms. This bridge allows us to attach essentially any Python sampling code to the well-tested and familiar \texttt{CosmoMC}, with little adaptation of the original Fortran code required. 

All three sampling methods return posterior distributions which are mutually consistent and also consistent with those given in the cosmology literature. We find that although optimised Metropolis-Hastings is in principle the fastest of the three methods for sampling probability distributions, in practice nested sampling and affine-invariant MCMC offer greater flexibility and robustness, requiring the adjustment of only 1 or 2 hyper-parameters. Metropolis-Hastings must be very precisely tuned to the target (posterior) distribution, and this is not possible even in principle when the distribution is non-Gaussian. 

Nested sampling and affine-invariant MCMC naturally find and explore parameter degeneracies with no {\it a priori} input from the user. The number of likelihood evaluations for ellipsoidal nested sampling scales exponentially with the number of free parameters, ruling this out as a practical sampling technique for models with many free parameters unless one uses parallel sampling from the ellipsoid to reduce the wall time of each step. Affine-invariant ensemble MCMC requires the highest number of likelihood evaluations on the cosmological data set, but this sampling prescription is highly parallelisable. Indeed, its potential performance on non-Gaussian, multi-modal and curving distributions means this technique is extremely powerful. Nested sampling, although principally developed as a tool for Bayesian evidence calculation, is shown to deliver low-noise estimates for posterior statistics for low computational cost. 

We note that extensions to single-ellipsoidal nested sampling extend the range of application of this sampling technique to multi-modal and curving distributions. We have shown that, for parameter estimation, nested sampling should be adopted over the popular Metropolis-Hastings sampling technique in many cases.

\section*{Acknowledgments}
RA would like to thank Phil Bull for his coding advice, and Graeme Addison for helpful discussions on affine-invariant sampling. RA is supported by an STFC Ph.D. studentship. JD acknowledges ERC grant 259505.

\bibliographystyle{mn2e}

\label{lastpage}
\clearpage

\end{document}